\documentstyle[12pt]{article}
\def\1{\mbox{I\hspace{-.15em}1}}

\def\b{\begin{equation}}
\def\e{\end{equation}}
\def\bee{\begin{enumerate}}
\def\eee{\end{enumerate}}

\title{Casimir Effect In Krein Space Quantization}
\author{H. Khosravi$^1$, M. Naseri$^{1,2}$, S. Rouhani $^{3}$ and M.V. Takook$^{1}$\thanks{e-mail:
takook@razi.ac.ir}, }
\date{\today}

\begin{document}

\maketitle {\it \centerline{$^1$  Department of Physics, Razi
University, Kermanshah, IRAN}  \centerline{\it $^2$ Islamic Azad
University, Kermanshah Branch, Kermanshah, IRAN} \centerline{\it
$^3$ Plasma Physics Research Centre, Islamic Azad University,}
\centerline{\it P.O.BOX 14835-157, Tehran, IRAN}}

\begin{abstract}

An explicit calculation of Casimir effect through an alternative
approach of field quantization \cite{gareta,ta4}, has been
presented in this paper. In this method, the auxiliary negative
norm states have been utilized, the modes of which do not interact
with the physical states or real physical world. Naturally these
modes cannot be affected by the physical boundary conditions.
Presence of negative norm states play the rule of an automatic
renormalization device for the theory.

\end{abstract}

\vspace{0.5cm} {\it Proposed PACS numbers}: 04.62.+v, 03.70+k,
11.10.Cd, 98.80.H \vspace{0.5cm}


\section{Introduction}

Due to appearance of infrared divergence in the two point function
for the minimally coupled scalar field in de Sitter space, a new
method of field quantization has been presented
\cite{gareta,dere}. The minimally coupled scalar field plays an
important role in the inflationary model as well as in the linear
quantum gravity \cite{ta,gagarerota}. Infrared divergence however
appears in the linear gravity and the minimally coupled scalar
field in de Sitter space with a great deal of similarity
\cite{ta}. In the case of the linear gravity, this divergence does
not manifest itself in the quadratic part of the effective action
in the one-loop approximation. This means that the pathological
behavior of the graviton propagator is gauge dependent and so
should not appear in an effective way as a physical quantity
\cite{anilto1}.

A covariant quantization of minimally coupled scalar field cannot
be constructed by positive norm states alone \cite{al}. To cope
with this problem the following method of quantization has been
utilized to achieve a naturally renormalized theory. It has been
proven that the use of the two sets of solutions (positive and
negative norms states) are an unavoidable feature for preservation
of $(1)$ causality (locality), $(2)$ covariance, and $(3)$
elimination of the infrared divergence for the minimally coupled
scalar field in de Sitter space \cite{gareta}. Preserving the
covariance principle and ignoring the positivity condition,
similar to Gupta-Bleuler quantization of the electrodynamic
equations in Minkowski space, we have performed the field
quantization in the Krein space, in combined Hilbert and
anti-Hilbert space \cite{mi}. Most interesting result of this
construction is the convergence of the Green's function at large
distances, which means that the infrared divergence is gauge
dependent \cite{gareta,ta3}. The ultraviolet divergence in the
stress tensor disappears as well, in other words the quantum free
scalar field in this method is automatically renormalized. The
effect of ``unphysical'' states (negative norm states) appears in
the above theory as a natural renormalization element.

It is important to note that by the use of this method, a natural
renormalization of the following problems, have been already
attained:
\begin{itemize}
    \item the massive free field in de Sitter space \cite{gareta},
    \item the graviton two point function in de Sitter space \cite{ta2},
    \item the one-loop
effective action for scalar field in a general curved space-time
\cite{ta5},
    \item tree level scattering amplitude of scalar field with one
    graviton exchange in de Sitter space \cite{rota1}.
\end{itemize}

Following above achievements and through the same new approach,
the Casimir force between two parallel plates in Minkowski space
has been calculated in this paper. In section $2$ we briefly
recall the Casimir force in the Hilbert space quantization.
Section $3$ is devoted to the calculation of the Casimir force in
the Krein space quantization in $2$ and $4$ dimensional space
time. Brief conclusion and outlook are given in final section.

\section{Casimir Force}

The Casimir effect is a small attractive force, which acts between
two parallel uncharged conducting plates. It is due to quantum
vacuum fluctuation of the field operator between two parallel
plates. For simplicity, we consider the two dimensional space time
and one component field in this section. We start with the
quantization of the scalar field $\phi(t,x)$. In two dimensions,
the scalar field equation, {\it i.e.} the Klein-Gordon equation,
is $(c=\hbar=1)$: \b \frac{\partial^{2}\phi(t,x)} {\partial
t^{2}}-\frac{\partial^{2}\phi(t,x)}{\partial x^{2}
}+m^{2}\phi(t,x)=0.\e Inner products, which defined the norms, is
defined by \cite{bida} \b
(\phi_1,\phi_2)=-i\int_{t=\mbox{cons.}}\phi_1(t,x)\stackrel{\leftrightarrow}
{\partial}_t\phi_2^*(t,x)dx.\e  Two sets of solutions of $(1)$ are
given by: \b u_p(k,x,t)=\frac{e^{i k x-iwt}}{\sqrt{(2\pi)2w}}
=\frac{e^{-ik.x}}{\sqrt{(2\pi)2w}},\;\;u_n(k,x,t)=\frac{e^{-i k
x+iwt}}{\sqrt{(2\pi)2w}} =\frac{e^{ik.x}}{\sqrt{(2\pi)2w}},\e
where $ w(k)=k^0=(k ^2+m^2)^{\frac{1}{2}} \geq 0$. These
$u(k,x,t)$ modes are orthonormalized by the following relations:
$$ (u_p(k,x,t),u_p(k',x,t))=\delta( k- k'),$$ $$
(u_n(k,x,t),u_n(k',x,t))=-\delta( k-k'),$$ \b
(u_p(k,x,t),u_n(k',x,t))=0.\e $u_p$ modes are positive norm states
and the $u_n$'s are negative norm states. By imposing the physical
boundary condition on the positive norm states,\b
u_p(k,0,t)=u_p(k,a,t)=0,\e we obtain \b
u_p(k_N,x,t)=(\frac{1}{a\omega_{N}})^{1/2}e^{- i\omega_{N}t}\\sin
k_{N}x,\e where:
\b\omega_{N}=(m^{2}+k_{N}^{2})^{1/2},\,k_{N}=\frac{N\pi}{a},\,N=1,2,3,..\;
 .\e This is a typical case where the Casimir effect arises. In
this case, the scalar product is:
 \b
(u_p(k_N,x,t),u_p(k_{N'},x,t))=\delta_{NN^{'}}.\e

The standard quantization of the field is performed by means of
the expansion \b
\phi(t,x)=\sum_{N}[a_{N}u_p(k_N,x,t)+a^{\dagger}_{N}u_p^*(k_N,x,t)],\e
where $a_{N},a^{\dagger}_{N}$  are the annihilation and creation
operators obeying the commutation relations
\b[a_{N},a^{\dagger}_{N^{'}}]=\delta_{NN^{'}},[a_{N},a_{N^{'}}]=[a^{\dagger}_{N},a^{\dagger}_{N^{'}}]=0.\e
The vacuum state in the presence of boundary condition is defined
by\b a_{N}|0\rangle=0.\e The vacuum energy of this state is
calculated at this stage. The operator of the energy density is
given by the $00$-component of the energy-momentum tensor of the
scalar field in the two-dimensional space-time \b
T_{00}=\frac{1}{2}\{[\partial_{t}\phi(t,x)]^{2}+[\partial_{x}\phi(t,x)]^{2}\}.\e
Substituting Eq. $(9)$ into Eq. $(12)$ accounting $(10)$ and
$(11)$ one easily obtains\b
\langle0|T_{00}|0\rangle=\frac{1}{2a}\sum_{N=1}^{\infty}\omega_{N}-\frac{m^{2}
}{2a}\sum_{N=1}^{\infty}\frac{\cos2k_{N}x}{\omega_{N}}.\e The
total vacuum energy of the interval $(0,a)$ is obtained by the
integration of $(13)$ as follow:\b
E_{0}(a)=\int_{0}^{a}\langle0|T_{00}|0\rangle
dx=\frac{1}{2}\sum_{N=1}^{\infty}\omega_{N}.\e The second,
oscillating term in the right-hand side of $(13)$ does not
contribute to this result.

The expression $(14)$ for the vacuum state energy of the quantized
field (between above boundaries) is generally the standard
starting point in the theory of the Casimir effect. Evidently the
quantity $E_{0}(a)$ is infinite. There are many regularization
procedures. Here we use one of the simplest methods, i.e., we
introduce an exponentially damping function
$exp(-\delta\omega_{N})$. In the limit $\delta\rightarrow0$ the
regularization is removed. For simplicity let us consider the
regularized vacuum energy of the above interval for a massless
field $(m = 0)$. In this case
 \b E_{0}(a,\delta)=\frac{1}{2}\sum_{N=1}^{\infty}\frac{\pi N}{a}\exp(-\frac{\delta \pi N}{a})
 =\frac{\pi }{8a}\sinh^{-2}\frac{\delta  \pi}{2a}.\e
In the limit of small $\delta$ one obtains; \b
E_{0}(a,\delta)=\frac{a}{2\pi \delta^{2}}+E(a)+O(\delta^{2}) ,
E(a)=-\frac{\pi}{24a},\e
 i.e., the vacuum energy is represented as a sum of a singular term and a
 finite contribution.

Let us compare this result with the corresponding result for the
unbounded axis. Here instead of (6) we have the positive frequency
solutions in the form of travelling waves $(3)$. The sum in the
field operator (9) is interpreted now as an integral with the
measure $\frac{dk}{2\pi}$, and the commutation relations contain
delta functions $\delta(k-k')$, instead of the Kronecker symbols.
Let us call the vacuum state defined by: \b a_{k}|0_{M}\rangle=0 ,
\e the Minkowski vacuum to underline the fact that it is defined
in free space without any boundary conditions. Repeating exactly
the same simple calculation which was performed for the above
interval, we obtain the divergent expression for the vacuum energy
density in Minkowski vacuum; \b
\langle0_{M}|T_{00}|0_{M}\rangle=\frac{1}{2\pi}\int_{0}^{\infty}\omega
dk,\e and for the total vacuum energy on the axis \b
E_{0M}(-\infty,+\infty)=\frac{1}{2\pi}\int_{0}^{\infty}\omega
dkL,\e where $L \rightarrow\infty$ is the normalization length.
Let us separate the interval $(0, a)$ of the whole axis whose
energy should be compared with $(14)$: \b
E_{0M}(a)=\frac{E_{0M}(-\infty,+\infty)}{L}a=\frac{
a}{2\pi}\int_{0}^{\infty}\omega dk.\e To calculate $(20)$ we use
the same regularization method as the above case, i.e., we
introduce the exponentially damping function under the integral;
\b E_{0M}(a)=\frac{ a}{2\pi}\int_{0}^{\infty}ke^{-\delta
k}dk=\frac{ a}{2\pi \delta^{2}}.\e  The obtained result coincides
with the first term in the right-hand side of $(16)$.
Consequently, the renormalized vacuum energy of the interval $(0,
a)$ in the presence of boundary conditions can be defined
as\cite{fu}: \b
E_{0}^{Ren}(a)=\lim_{\delta\rightarrow0}[E_{0}(a,\delta)-E_{0M}(a,\delta)]=-\frac{\pi
}{24a}.\e

This leads us to calculation of following attractive force between
the above boundaries: \b F(a)=-\frac{\pi}{24 a^{2}}\e In this
case, the renormalization corresponds to removing the vacuum
energy of the unbounded space from the total energies of
corresponding bounded case. The renormalized energy $E(a)$
monotonically decreases as the boundary points approach each
other. This points to the presence of an attractive force between
the conducting planes.

\section{Casimir force in Krein space quantization}

In this section we calculate zero point energy and casimir force
in Krein space quantization thoroughly. In the previous paper
\cite{ta4}, we present the free field operator in the Krein space
quantization \b \phi(t,x)=\phi_p(t,x)+\phi_n(t,x),\e where
$$ \phi_p(t,x)=\int dk [a(k)u_p(k,x,t)+a^{\dag}(
k)u_p^*(k,x,t)],$$ $$ \phi_n(t,x)=\int dk [b(
k)u_n(k,x,t)+b^{\dag}(k)u_n^*(k,x,t)],$$ and $a(k)$ and $b(k)$ are
two independent operators. Creation and annihilation operators are
constrained to obey the following commutation rules \b [a(
k),a(k')]=0,\;\;[a^{\dag}( k), a^{\dag}(k')]=0,\;\;, [a(
k),a^{\dag}(k')]=\delta(k-k') ,\e \b [b(k),b(
k')]=0,\;\;[b^{\dag}( k), b^{\dag}( k')]=0,\;\;, [b( k),b^{\dag}(
k')]=-\delta( k-k') ,\e
 \b [a( k),b(k')]=0,\;\;[a^{\dag}(k), b^{\dag}( k')]=0,\;\;, [a
(k),b^{\dag}(k')]=0,\;\;[a^{\dag}( k),b(k')]=0 .\e The vacuum
state $\mid \Omega>$ is then defined by \b a^{\dag}(k)\mid
\Omega>= \mid 1_{ k}>;\;\;a(k)\mid \Omega>=0, \e \b b^{\dag}(
k)\mid \Omega>= \mid \bar1_{k}>;\;\;b( k)\mid \Omega>=0, \e \b b(
k)\mid 1_{ k}
>=0;\;\; a(k)\mid \bar1_{k} >=0, \e
where $\mid 1_{k}> $ is called a one particle state and $\mid
\bar1_{k}>$ is called a one ``unparticle state''.

By imposing the physical boundary condition on the field operator,
only the positive norm states are affected. The negative modes do
not interact with the physical states or real physical world, thus
they can not be affected by the physical boundary conditions as
well. In this case, the field operator is defined as: \b
\phi(t,x)=\sum_{N}[a_{N}u_p(k_N,x,t)+a^{\dagger}_{N}u_p^*(k_N,x,t)]+
\int dk [b( k)u_n(k,x,t)+b^{\dag}(k)u_n^*(k,x,t)].\e Substituting
the above field operator in $(12)$ and using Eqs $(28)$ and
$(29)$, one easily obtains:
\b\langle\Omega|T_{00}^{Kre}|\Omega\rangle=\frac{1}{2a}\sum_{N=1}^{\infty}\omega_{N}-\frac{m^{2}}{2a}\
\sum_{N=1}^{\infty}\frac{\cos2k_{N}x}{\omega_{N}}-\frac{1}{2\pi}\int_{0}^{\infty}\omega
dk. \e The total vacuum energy for the interval $(0,a)$ is
obtained by the integration of $(32)$ as follow:\b
E_{0}^{Kre}(a)=\int_{0}^{a}\langle\Omega|T_{00}^{Kre}|\Omega\rangle
dx=-\frac{\pi}{24a},\e which is exactly the previous result. Then
the attractive Casimir force between the conducting planes in the
Krein space quantization is then \b
F(a)=-\frac{\partial(E_{0}^{Kre}(a))}{\partial
a}=-\frac{\pi}{24}\frac{1}{ a^{2}}\;.\e

The Casimir force for the Electromagnetic field in $4$-dimensional
space time, {\it i.e.} a real physical case, in the Krein space
quantization has been calculated as well. It resulted in the
following Casimir force \cite{itzu}\b
F(a)=-\frac{\partial(E_{0}^{Kre}(a))}{\partial
a}=-\frac{\pi^2}{240 } \frac{1}{a^{4}}\;. \e Due to its similarity
with the above $2$-dimensional case it was not necessary to
present the explicit calculation. It is important to note that
once again the natural renormalization of the theory has been
established.

\section{Conclusion}

The negative frequency solutions of the field equation are needed
for the covariant quantization in the minimally coupled scalar
field in de Sitter space. Contrary to the Minkowski space, the
elimination of de Sitter negative norms in this case breaks the de
Sitter invariance. In other words for restoring of the de Sitter
invariance, one needs to take into account the negative norm
states {\it i.e.} the Krein space  quantization. This provides a
natural tool for renormalization of the theory \cite{gareta}. In
the present paper, Casimir force in Minkowski space-time, has been
calculated through the Krein space quantization. Once again it is
found that the theory is automatically renormalized.\vspace{0.5cm}

\noindent {\bf{Acknowlegements}}: The author would like to thank
S. Teymourpoor for its interest in this work.

\end{document}